\documentclass[prb,aps,twocolumn]{revtex4}
\begin{document}
\title{Comment on ``Quasisaddles as relevant points of the potential energy 
surface in the dynamics of supercooled liquids'' [J.\ Chem.\ Phys.\ {\bf 116}, 10297 (2002); cond-mat/0203301]}
\author{Jonathan P. K. Doye, and David J. Wales}
\affiliation{University Chemical Laboratory, Lensfield Road, Cambridge CB2 1EW,
United Kingdom}
\maketitle
Following theoretical work that emphasized the importance of saddle points 
on a potential energy surface (PES) in the dynamics of supercooled liquids,\cite{Cavagna01}
two papers appeared proposing a mapping of
instantaneous configurations onto the 
underlying saddle points of the PES.\cite{Angelani00,Broderix00} 
In this mapping the instantaneous configurations are starting
points for minimizations on the function $W=|\nabla V|^2$, 
where $V$ is the potential energy. 
The motivation for this choice was that the stationary points of
the PES, of whatever index, 
correspond to absolute minima of $W$ with $W=0$.
However, we subsequently pointed out that
the vast majority of the minima of $W$ do not 
correspond to stationary points of the PES, but have $W>0$.\cite{Doye02a}
This result was recently confirmed by the authors of Ref.~\onlinecite{Angelani00}
after reanalysing their data.\cite{Angelani02} 

The minima of $W$ must satisfy $\nabla W=2 {\bf H}\,{\bf g}={\bf 0}$ where 
${\bf H}$ is the Hessian (the second derivative
matrix of the potential energy) and ${\bf g}$ is the gradient, $\nabla V$.
There are two types of solution to this equation. First, those where ${\bf g}=0$, 
which correspond to true stationary 
points of the PES, and second, those where ${\bf g}$ is an eigenvector of the Hessian with zero eigenvalue. 
These minima have $W>0$ and have a point of inflection in the direction of the gradient on the PES,
and we will continue to refer to them as non-stationary points (NSP's) in order
to differentiate them from the first set of minima.

In Ref.~\onlinecite{Angelani02} Angelani {\it et al.\/} now argue that 
it is reasonable to consider the NSP's as `quasi-saddles':
`If this number [the number of inflection directions at the minima of $W$] is small
one can think of the minima of W as a ``quasi'' saddle point (QSP), 
as they are {\em true} saddle points in the subspace orthogonal to the small number of 
inflection directions. One can conjecture that the properties of a QSP are very similar to 
that of a true saddle point associate (sic) to an instantaneous configuration'.
In this comment we suggest that these properties
simply follow from the form of the solutions to $\nabla W=0$, and
do not really indicate any saddle-like character. 

We first note that NSP's are expected to have 
exactly one zero Hessian eigenvalue (in addition to the three zeros corresponding 
to translation of the whole system), unless there is some symmetry present in the system giving
rise to degenerate eigenvalues, or an unexpected accidental degeneracy, both of which are very unlikely
in the present case. 
This analytic result is confirmed by an analysis of 
the 8,000 stationary points of $W$ of the 256-atom system in our database.\cite{Doye02a}
Only eight of the 6,920 NSP's have a non-zero Hessian eigenvalue with magnitude less than $0.001$. 
Using the convergence criteria described in Ref.~\onlinecite{Doye02a} 
the average magnitude of the zero eigenvalue for the NSP's is 
$6\times10^{-5}$,
and the average magnitude of the smallest non-zero eigenvalue over all 8,000 points is $0.64$.\cite{units}
We therefore suggest that the results in Ref.~\onlinecite{Angelani02}, where the
average number of directions with
zero curvature varies from between 1.1 and 4.4 (depending on temperature),
are due to incomplete minimization, and that the tolerance used to assign the zero eigenvalues
in Ref.~\onlinecite{Angelani02} is not tight enough.
Instead, the number of extra zero Hessian eigenvalues should be precisely one, 
independent of temperature.

Secondly, we note that at an NSP the inflection direction is parallel to the gradient. Hence the gradient 
in the sub-space orthogonal to the inflection direction is zero, and the NSP
is a stationary point in this subspace. However, any point in configuration space is
a stationary point in the subspace orthogonal to the gradient, and so this result 
does not imply any additional saddle-like properties.

The authors of Ref.\ \onlinecite{Angelani02} introduce a further justification for the use of the
term `quasisaddle' in a footnote: `most of the QSP found are ``almost'' saddles in the
sense that $W$ is almost zero numerically'. It is not surprising that the magnitude
of the gradient at the NSP's is significantly smaller than at the instantaneous configuration, 
since this is what is being minimized. However, analysing our database shows that 
$W$ is not almost zero numerically. 
The average value of $W$ at a NSP varies between 0.28 and 3.25 depending on temperature,
compared to a value for the true stationary points of $2\times 10^{-8}$, 
i.e. numerically zero consistent with our convergence criteria.

We therefore disagree with the arguments used in Ref.~\onlinecite{Angelani02}
to infer saddle-like properties for NSP's. 
The most interesting results of Angelani {\it et al.\/} do not depend
upon this classification---the observation that the number of negative eigenvalues of the NSP's extrapolates to zero
at the mode-coupling temperature, and is strongly correlated with the
magnitude of the diffusion coefficient, is still intriguing.
However, the interpretation of these results depends upon a 
correct understanding of the nature of the NSP's.


\begin{thebibliography}{1}
\bibitem{units}
As in Ref.\ \onlinecite{Doye02a} we use $\epsilon_{AA}$ and $\sigma_{AA}$ as
  the units of energy and distance.
\end{thebibliography}
\end{document}